\documentclass[11pt]{article}
\usepackage{graphicx}
\usepackage{fancyhdr}
\usepackage{makeidx}
\usepackage{amssymb,amsmath}
\usepackage{enumerate}
\newcommand{\bd}{\begin{document}}
\newcommand{\ed}{\end{document}}
\newcommand{\bc}{\begin{center}}
\newcommand{\ec}{\end{center}}
\newcommand{\vs}{\vspace}
\newcommand{\hs}{\hspace}
\newcommand{\beq}{\begin{equation}}
\newcommand{\eeq}{\end{equation}}
\newcommand{\beqs}{\begin{eqn*}}
\newcommand{\eeqs}{\end{eqn*}}
\newcommand{\bq}{\begin{quote}}
\newcommand{\eq}{\end{quote}}
\newcommand{\lb}{\linebreak}
\newcommand{\mb}{\makebox}
\newcommand{\fb}{\framebox}
\newcommand{\mc}{\multicolumn}
\newcommand{\ben}{\begin{enumerate}}
\newcommand{\een}{\end{enumerate}}
\newcommand{\bit}{\begin{itemize}}
\newcommand{\eit}{\end{itemize}}
\newcommand{\ov}{\overline}
\newcommand{\un}{\underline}
\newcommand{\lt}{\left}
\newcommand{\rt}{\right}
\newcommand{\ba}{\begin{array}}
\newcommand{\ea}{\end{array}}
\newcommand{\beqa}{\begin{eqnarray}}
\newcommand{\eeqa}{\end{eqnarray}}
\newcommand{\beqas}{\begin{eqnarray*}}
\newcommand{\eeqas}{\end{eqnarray*}}
\newcommand{\bfg}{\begin{figure}}
\newcommand{\efg}{\end{figure}}
\newcommand{\pad}{\partial}
\newcommand{\nn}{\nonumber}
\newcommand{\la}{\leftarrow}
\newcommand{\ra}{\rightarrow}
\newcommand{\lgla}{\longleftarrow}
\newcommand{\lgra}{\longrightarrow}
\newcommand{\La}{\Leftarrow}
\newcommand{\Ra}{\Rightarrow}
\newcommand{\Lra}{\Leftrightarrow}
\newcommand{\Lgla}{\Longleftarrow}
\newcommand{\Lgra}{\Longrightarrow}
\renewcommand{\a}{\alpha}
\renewcommand{\b}{\beta}
\newcommand{\g}{\gamma}
\newcommand{\G}{\Gamma}
\renewcommand{\d}{\delta}
\newcommand{\D}{\Delta}
\newcommand{\e}{\epsilon}
\newcommand{\eps}{\epsilon}
\newcommand{\s}{\sigma}
\renewcommand{\l}{\lamda}
\newcommand{\m}{\mu}
\newcommand{\n}{\nu}
\renewcommand{\S}{\Sigma}
\newcommand{\p}{\pi}
\newcommand{\om}{\omega}
\newcommand{\Om}{\Omega}
\newcommand{\tri}{\triangle}
\newcommand{\ti}{\times}
\newcommand{\f}{\frac}
\newcommand{\ds}{\displaystyle}
\newcommand{\bm}[1]{\mb{{\boldmath $#1$}}}
\newcommand{\alter}[2]{\lt\{ \ba{ll}#1 \\ #2 \ea \rt.}
\newcommand{\alt}[4]{\lt\{ \ba{ll}#1 & \mb{if \, \,}#2 \\ #3 & \mb{}#4 \ea
    \rt.}
\newcommand{\altn}[4]{\lt\{ \ba{rl}#1 & \mb{if \, \,}#2 \\ #3 & \mb{}#4 \ea
    \rt.}
\newcommand{\altif}[4]{\lt\{ \ba{ll}#1 & \mb{if \, \,}#2 \\ #3 &
\mb{if \, \,}#4 \ea \rt.}
\newcommand{\altnif}[4]{\lt\{ \ba{rl}#1 & \mb{if \, \,}#2 \\ #3 &
\mb{if \, \,}#4 \ea \rt.}
\newcounter{algc}
\newcounter{romc}
\newcounter{Alphc}
\newcommand{\bl}{\begin{list}{{\it Step} ~\arabic{algc}~:} {\usecounter{algc}
                \setlength{\topsep}{0pt} \setlength{\itemsep}{0pt}}}
\newcommand{\el}{\end{list}}
\newcommand{\blr}{\begin{list}{~\roman{romc}~:} {\usecounter{romc}
                \setlength{\topsep}{0pt} \setlength{\itemsep}{0pt}}}
\newcommand{\elr}{\end{list}}
\newcommand{\bla}{\begin{list}{~\Alph{Alphc}~:} {\usecounter{Alphc}
                \setlength{\topsep}{0pt} \setlength{\itemsep}{0pt}}}
\newcommand{\ela}{\end{list}}

\newtheorem{theorem}{Theorem}
\begin{document}
\title{Intrinsic Limits of Subthreshold Slope in Biased Bilayer Graphene Transistor}
\author{Kausik Majumdar\thanks{Corresponding author: Kausik Majumdar, Email: kausik@ece.iisc.ernet.in},
Kota V. R. M. Murali$^{\dag}$, Navakanta Bhat and Yu-Ming Lin$^{\dag\dag}$\\
Department of Electrical Communication Engineering and \\Center
of Excellence in Nanoelectronics,\\ Indian Institute of Science,
Bangalore-560012, India.\\
$^{\dag}$IBM Semiconductor Research and Development Center, Bangalore -
560045, India.\\
$^{\dag\dag}$IBM T. J. Watson Research Center, Yorktown
Heights, NY 10598, USA.\\}
\date{}
\maketitle {\abstract \textbf{In this work, we investigate the
intrinsic limits of subthreshold slope in a dual gated bilayer
graphene transistor using a coupled self-consistent
Poisson-bandstructure solver. We benchmark the solver by matching
the bias dependent bandgap results obtained from the solver against
published experimental data. We show that the intrinsic bias
dependence of the electronic structure and the self-consistent
electrostatics limit the subthreshold slope obtained in such a
transistor well above the Boltzmann limit of 60mV/decade at room
temperature, but much below the results experimentally shown till
date, indicating room for technological improvement of bilayer
graphene.}}

\newpage
The excellent transport properties of single and multi-layer
graphene hold promise to build ultra-fast transistors
with excellent on state characteristics \cite{rs98}-\cite{jbo08}.
However, the lack of significant bandgap in such systems has been
one of the major roadblocks to achieve low off state current and
hence high on/off current ratio \cite{mcl07}-\cite{jbo08}.
Recently, it has been found, both theoretically and
experimentally, that a bandgap can be opened up in a bilayer
graphene (BLG) using an external bias \cite{jbo08}-\cite{fxia10}. A
recent experiment shows that a bandgap as large as $\sim$0.3eV can
be created in a BLG depending on the external bias \cite{yz09}. This
relatively large bandgap is promising to obtain low off state
current, hence an improved on to off current ratio and subthreshold
slope \cite{fxia10}.

The aim of this paper is to discuss the intrinsic limits of the
subthreshold slope in a BLG metal-oxide-semiconductor field effect
transistor (MOSFET). We assume a long channel transistor with an
ideal gate dielectric interface. To achieve this, we develop a
self-consistent Poisson-bandstructure solver for a dual gated BLG
device. We have simulated the magnitude of the bias dependent
bandgap tuning in a BLG system which matches closely with the
published experimental data \cite{yz09} validating the solver. We
also show that the gate bias dependent electronic structure and the
self-consistent electrostatics play a central role in determining
the intrinsic limits of the subthreshold slope in a BLG transistor,
which remains well above the room temperature Boltzmann limit of 60
mV/decade.

A schematic of the BLG device configuration is shown in Fig.
\ref{fig:Eg}(a) where a BLG is sandwiched between a top and a bottom
gate stack. Each gate stack consists of a gate dielectric with
an equivalent oxide thickness (EOT) of 1nm
and a gate contact metal.
We assume perfect dielectric-BLG interface and zero flatband voltage.
The voltage at the top gate ($V_t$) and the bottom gate
($V_b$) can be varied independently. The self-consistent electronic
structure of this device configuration is determined by the
bandstructure of the BLG using Tight Binding Method
(\cite{rs98,evc08,jn08}) coupled with the 1-D Poisson equation. Taking the center of
the two graphene layers at $z=0$, the charge density is given by
$\rho(z)=qn(z)$ where $q$ is the electronic charge and
$n(z)$ is obtained as a subtraction between the hole ($n_h(z)$)
and electron ($n_e(z)$) carrier density:
\beq\label{eq:n}
n(z)=2\left[\sum_{i,\bar{k}}(1-f(E_i(\bar{k})))|\psi_i^{\bar{k}}(z)|^2
-\sum_{j,\bar{k}}f(E_j(\bar{k}))|\psi_j^{\bar{k}}(z)|^2\right]
\eeq
where $f(E_{i,j}(k))$ is
the Fermi-Dirac probability of the state $(\{i,j\},\bar{k})$ at
temperature $T$. $i$ and $j$ are the valence and conduction band indices
respectively.  The energy eigenvalues $E_{i,j}(\bar{k})$
are obtained from the tight binding bandstructure taking
only $p_z$ orbital into account, with an intra-layer overlap integral
($S$) of 0.129, the intra-layer hopping $t$ as $-$3.033eV and
inter-layer hopping $t_\perp$ as $-$0.365eV \cite{rs98}. To obtain
the wavefunction $\psi_i^{\bar{k}}(z)$, we assume normalized
Gaussian orbital as the basis function. The wavefunctions
are set to zero at the BLG-dielectric interfaces.

We validate the above self-consistent method by comparing the bias
dependent bandgaps of BLG experimentally obtained in \cite{yz09}. In
this case, $V_t$ is set to $-V_b$ which breaks the inversion
symmetry of the BLG opening a bandgap \cite{jbo08, to06}. The
results are shown in Fig. \ref{fig:Eg}(b) where the bandgap ($E_g$)
is plotted as a function of the average displacement vector $D$,
defined in the same way as in \cite{yz09}.
Here the relative permittivity ($\eps_r$) of BLG is varied
as a parameter to find that $\eps_r$=1.8 provides the best fit to the
experimental data and the same value is used in the Poisson equation to generate the
results discussed in the rest of the paper.
Fig. \ref{fig:Eg}(c) shows the carrier density profiles along $z$ at
$D$=3V/nm. This clearly shows a strong charge polarization in the bilayer,
though the system as a whole remains almost intrinsic for
$V_t=-V_b$.

The recent development of significant tunability of bandgap in a
bilayer graphene (\cite{to06}-\cite{jn08}) brings the possibility
of significant reduction of off current in a BLG transistor. In the
following, we show that the bias dependent electronic structure and
the corresponding self-consistent electrostatics play a major
role in determining the intrinsic limits of the subthreshold slopes
and the on/off current ratio in an ideal long channel BLG MOSFET.

Before discussing the self-consistent results, we first show that
even in absence of the screening effect,
the intrinsic bias dependence of the electronic structure in BLG
can be a major cause for subthreshold slope
degradation. This arises from the dependence of the rate of change of
electron and hole barrier height ($B_{e,h}$) with gate bias.
This rate, ideally, should be as high as
possible, with the maximum possible value being $1$, which
corresponds to a subthreshold slope of 60mV/decade. This is obtained
by finding the roots
$\lambda$ of the secular equation of the bilayer Hamiltonian matrix
(with $S=0$):
\setlength{\arraycolsep}{-2.8em}
\begin{eqnarray}\label{eq:ns}
\left[(V_t-\lambda)^2-t^2|s_k|^2\right]\left[(V_b-\lambda)^2-t^2|s_k|^2\right] \nonumber \\
&&+ t_{\perp}^2(V_t-\lambda)(V_b-\lambda)=0
\end{eqnarray}
\setlength{\arraycolsep}{5pt}
where $s_k=1+e^{i\bar k.\bar{a_1}}+e^{i\bar k.\bar{a_2}}$, $\bar{a_1}$ and $\bar{a_2}$
are lattice vectors in graphene. In Fig. \ref{fig:Eg}(d), we plot
the rate of change of $B_e$ for different directions
in the $(V_t,V_b)$ space in polar coordinates with
$\theta=tan^{-1}(V_t/V_b)$. This clearly shows that the rate is
maximum (=1) along $V_t=V_b+V_0$, which is essentially the constant
bandgap locus. As we deviate from this direction, the rate degrades
from $1$ with minimum along $V_t=-V_b$, leading to the poor
subthreshold slope along anti-symmetric bias case. As observed from
Fig. \ref{fig:Eg}(d), this is a fairly generic result for a large
variety of points in the subthreshold region.

We now consider a chemically undoped long channel BLG MOSFET with
metal source and drain having low drain bias. The
drain current is assumed to be completely dominated by the
thermionic carriers, neglecting tunneling. Under these
approximations, the electron barrier height $B_e$ is plotted in Fig. \ref{fig:ss}(a) as a
function of $V_t$ and $V_b$.
In the same
plot, we also show a number of possible paths $p$ from transistor off
state (labeled X$_p$ points) to on state (labeled Y$_p$ points).

Due to the ambipolar nature of the device, it is important to select the appropriate off
state for the device. The total integrated carrier density
($N=N_e+N_h$) is shown in Fig. \ref{fig:ss}(b) in the ($V_t,V_b$)
space to help choose the appropriate set of off state points.
In Fig. \ref{fig:ss}(a), X$_{1,4}$
corresponds to the point of maximum bandgap in that range. X$_2$
corresponds to the maximum electron barrier height though it reduces
the hole barrier height, hence increasing $N$. X$_3$ is an intermediate point
with symmetric electron and hole barrier heights. We now compare the
characteristics among these off to on paths shown in Fig.
\ref{fig:ss}(a), namely (1): $V_b$ is fixed at 1V and $V_t$ is
varied from $-1$V to 1V,
(2): $V_b$ is fixed at 0.5V and $V_t$ is varied from $-$1V to 1V,
(3): $V_b=V_t+1$V, and $V_t$ is varied from $-1$V to 0V, which is an
almost constant bandgap operation (with $E_g\approx 0.17$eV), (4):
$V_b=-V_t$ and $V_t$ is varied from $-1$V to 0V, which is the
maximum bandgap tunability path. It should be noted that (1) and (2)
correspond to single gate operation with back gate at fixed bias. On
the other hand, (3) and (4) correspond to double gate operation,
although unconventional, with top and bottom gates at different
biases. The conventional double gate operation with $V_t=V_b$ is of
little significance due to zero bandgap. Fig. \ref{fig:ss}(c)
shows $N$ as a function of $V_t$ in these four cases. It is clearly
observed that the best on/off ratio obtained is $\sim$100.
Also, path (1) and (4) will have lower off current than (2) and (3), which
is also expected from the Fig. \ref{fig:ss}(b). Path (4) is found to
have low on/off ratio due to the poor tunability of barrier
heights (though bandgap tunability is maximum) in the anti-symmetric
case.

To get insights into the above,
we find the drain current in a long channel transistor as $I_d=qN\tau$,
where $\tau$ is the transport factor. In a MOSFET configuration,
$\tau$ is a weak function of gate voltage. Hence, the subthreshold
slope is $SS=ln(10)\frac{\partial V_t}{\partial ln(I_d)}\approx
ln(10)\frac{\partial V_t}{\partial ln(N)}$. In Fig.
\ref{fig:ss}(d), the computed subthreshold slopes are plotted for
the paths (1) to (4) with varying $V_t$. Note that, unlike conventional
MOSFET, the computed subthreshold slope is not independent of gate bias
in the BLG transistor.

Analytically, we can
convert the summation used in Eq. \ref{eq:n} into an integral
to obtain $N$:
\beq N=2\left[\int_{-\infty}^{E_v} D_v(E)(1-f(E))dE +
\int_{E_c}^\infty D_c(E)f(E)dE\right] \eeq where only one band is
assumed to contribute in the conduction band (CB) and the valence
band (VB). $E_{v,c}$ are the VB$_{max}$ and CB$_{min}$ obtained from
the self-consistent electrostatics. $D_{v,c}$ are the 2-D DOS of VB
and CB respectively, and we replace them by an average DOS $D$ using
the fact that close to band edge, $D_{v,c}$ change much slowly as
compared to Fermi function \cite{evc08}. Utilizing CB-VB symmetry,
we find, \beq N = 2k_BTD\ln\left[(1+e^{-B_e'})(1+e^{-B_h'})\right]
\eeq where $B_e'=\frac{B_e}{k_BT}=\frac{E_c-\mu}{k_BT}$ and
$B_h=\frac{B_h}{k_BT}=\frac{\mu-E_v}{k_BT}$. As $D$ is weakly
dependent on $V_t$ \cite{evc08}, we obtain \beq\label{eq:ss} SS=
-\ln(10)\frac{k_BT\kappa\ln(\kappa)}{\left[e^{-B_e'}\frac{\partial
B_e}{\partial V_t} + e^{-B_h'}\frac{\partial B_h}{\partial V_t} +
e^{-E_g'}\frac{\partial E_g}{\partial V_t}\right]} \eeq where
$\kappa=(1+e^{-B_e'})(1+e^{-B_h'})$ and
$E_g'=\frac{E_g}{k_BT}=\frac{B_e+B_h}{k_BT}$. In a conventional Si
MOSFET, in the subthreshold region, $B_e',B_h'>>0$, $E_g$ is
constant and no gate field screening occurs due to small carrier
density ($|\frac{\partial B_{e,h}}{q\partial V_t}|$=1). Under these
conditions, it is straight forward to show that $SS$ reduces to
$\ln(10)(\frac{k_BT}{q})$ (=60mV/decade at $T=300$K). Unfortunately,
for bilayer graphene, these three conditions are never satisfied
simultaneously which degrades the subthreshold slope. To open up a
bandgap in BLG, unlike conventional semiconductor, we need to apply
significantly large fields at the two gates. This in turn causes a
large number of carriers to be present in the device, even in
subthreshold (though it is possible that the overall device is close
to neutral, like in the case of $V_t=-V_b$, Fig. \ref{fig:Eg}(c)).
These carriers cause a strong screening of the gate field and hence,
a reduction in the derivatives in the denominator of Eq.
\ref{eq:ss}. Thus, even if we create a significant bandgap using
vertical field, the carriers thus generated degrade the subthreshold
slope.

In the anti-symmetric bias case (path (4)), using $B_e=B_h=\frac{E_g}{2}$,
we find from Eq. \ref{eq:ss} that the
$SS$ limit reduces to $60/|\frac{\partial B_{e}}{q\partial V_t}|$.
As $|\frac{\partial B_{e}}{q\partial V_t}|$ is intrinsically small in this direction
(see Fig. \ref{fig:Eg}(d)),
$SS$ degrades significantly. On the other hand,
along the constant bandgap paths (e.g. path 3),
the $SS$ limit again reduces to $60/|\frac{\partial B_{e}}{q\partial V_t}|$.
However, in this case, the strong screening due to large number of carriers
reduces the denominator, limiting $SS$ much above Boltzmann limit.
Finally, we have numerically performed a
global search in the ($V_t,V_b$) space with $|V_{t,b}|\leq1$V,
which shows that the best $SS$ for
all possible constant bandgap paths is around 200mV/decade, whereas
along all possible constant $V_b$ paths is around 275mV/decade.

Thus, we find that along the constant bandgap loci, the
fundamental limit of $SS$ arising from the bias dependent electronic
structure is 60mV/decade (see Fig. \ref{fig:Eg}(d)), but the
screening due to the presence of large number of carriers arising
from the unavoidable gate field degrades the subthreshold slope
significantly. However, as the operating condition moves away from
this direction, both the electronic structure as well as the
screening effect play major roles to degrade $SS$. Nonetheless, the
shown subthreshold slope limits are significantly less than the most recent
experimental data reported \cite{fxia10}
where the extracted value of the minimum subthreshold slope is
$\sim$550 mV/decade. This clearly shows that there is room for
significant technological improvement in bilayer graphene, including
efforts in improving transport derived bandgap, effective oxide thickness
and choosing the optimized path from on state to off state.

In conclusion, in this work, we have developed a coupled
self-consistent Poisson-bandstructure solver for dual gated bilayer
graphene. The bias dependent bandgap results obtained from the
solver have been benchmarked against published experimental data.
Finally, it has been shown that the bias dependent electronic
structure and the self-consistent electrostatics limit the
subthreshold slope obtained in such a transistor well above the
Boltzmann limit of 60mV/decade, but again much below the results
experimentally shown till date.\\

Two of the authors (K.M. and N.B.) would like to sincerely acknowledge
the support from the Ministry of Communication and Information Technology (MCIT),
Govt. of India, and the Department of Science and Technology (DST),
Govt. of India.

\newpage
\bfg[h]
\bc
\vs{-0.5in}
\includegraphics[scale=0.6]{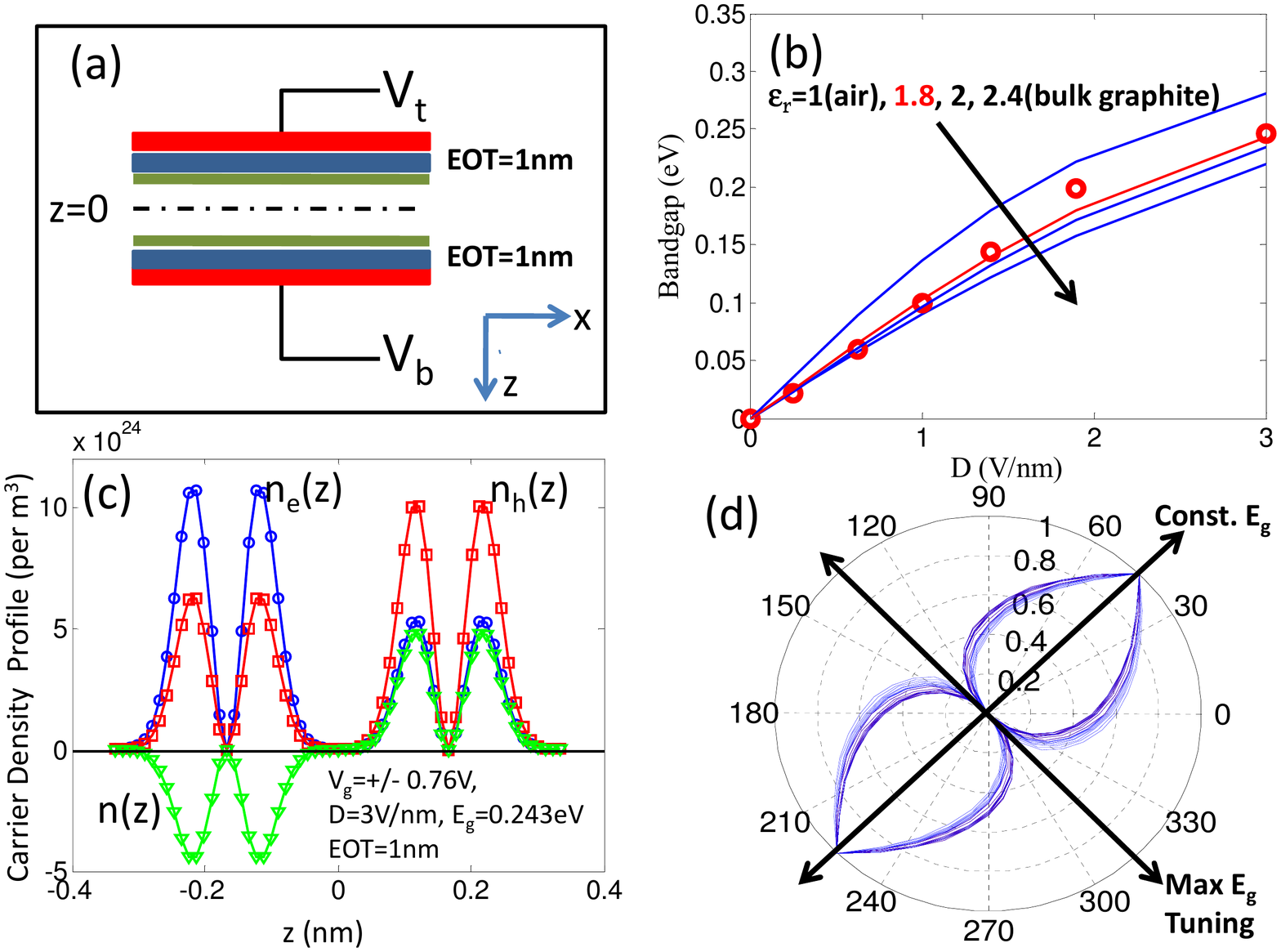}
\caption{(a): Schematic
of a bilayer graphene (BLG) sandwiched between two gate stacks. (b):
Verification of numerically obtained bandgap results with
experimental data in \cite{yz09}, shown as red dots. (c): Carrier
density profiles from top layer to bottom layer. (d): Polar plot of
the rate of change of $B_e$ (normalized by $q$) along different
directions using Eq. \ref{eq:ns} (with angle
$\theta=tan^{-1}(V_t/V_b)$) in the ($V_t,V_b$) space for two sets of
subthreshold points: i) $V_t$=$-1$V, $V_b\in[0V, 1V]$ and ii)
$V_t$=$-0.5$V, $V_b\in[0V, 0.5V]$. All the curves show similar
directional dependence.}\label{fig:Eg} \ec \efg

\newpage
\bfg[h]
\bc
\vs{-0.5in}
\includegraphics[scale=0.6]{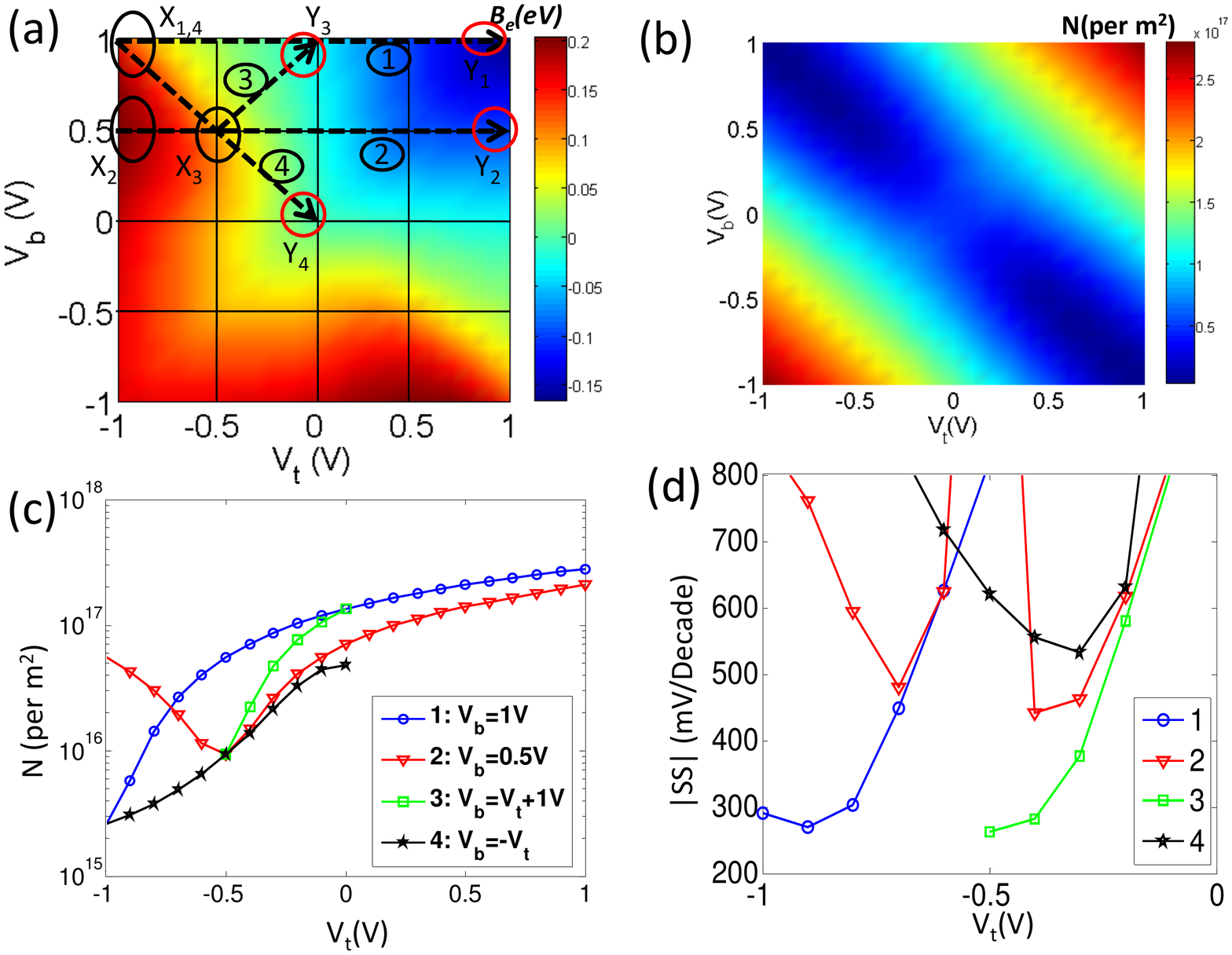}
\caption{(a): Electron Barrier height $B_e$ as
a function of $V_t$ and $V_b$. The four different off to on state
arrow paths are numbered as (1) to (4). (b): Integrated carrier density
($N$) as a function of $V_t$ and $V_b$. (c): N along the different
off to on paths (d): The absolute values of the computed subthreshold slopes for the
paths (1)-(4) without any non-ideality.}\label{fig:ss}
\ec
\efg
\end{document}